\documentclass[12pt,useAMS,usegraphicx]{mn2e}

\begin{document}
\newcommand{\ang}{\rm \AA}
\newcommand{\msun}{M$_\odot$}
\newcommand{\lsun}{L$_\odot$}
\newcommand{\days}{$d$}
\newcommand{\degree}{$^\circ$}
\newcommand{\ud}{{\rm d}}
\newcommand{\as}[2]{$#1''\,\hspace{-1.7mm}.\hspace{.0mm}#2$}
\newcommand{\am}[2]{$#1'\,\hspace{-1.7mm}.\hspace{.0mm}#2$}
\newcommand{\ad}[2]{$#1^{\circ}\,\hspace{-1.7mm}.\hspace{.0mm}#2$}
\newcommand{\lsim}{~\rlap{$<$}{\lower 1.0ex\hbox{$\sim$}}}
\newcommand{\gsim}{~\rlap{$>$}{\lower 1.0ex\hbox{$\sim$}}}
\newcommand{\HA}{H$\alpha$}
\newcommand{\HII}{\mbox{H\,{\sc ii}}}
\newcommand{\kms}{\mbox{km s$^{-1}$}}
\newcommand{\HI}{\mbox{H\,{\sc i}}}
\newcommand{\NaI}{\mbox{Na\,{\sc i}}}
\newcommand{\KI}{\mbox{K\,{\sc i}}}
\newcommand{\nan}{Nan\c{c}ay}
\newcommand{\galex}{{\it GALEX}}
\newcommand{\jks}{Jy~km~s$^{-1}$}

\title[A Neutral Hydrogen Shell Surrounding IRC+10216]{Discovery 
of a Shell of Neutral Atomic Hydrogen Surrounding 
the Carbon Star IRC+10216}

\author[Matthews et al.]{L. D. Matthews$^{1}$,  E. G\'erard$^{2}$, 
T. Le~Bertre$^{3}$ \\
$^{1}$Massachusetts Institute of Technology 
Haystack Observatory, Off Route 40, Westford, MA
  01886 USA\\
$^{2}$GEPI, UMR 8111, Observatoire de Paris, 5 Place J.
Janssen, F-92195 Meudon Cedex, France\\
$^{3}$LERMA, UMR 8112, Observatoire de Paris, 61 av.
de l'Observatoire, F-75014 Paris, France}

\maketitle
\begin{abstract}
We have used
the Robert C. Byrd Green Bank Telescope to perform the most
sensitive search to date for neutral atomic hydrogen (\HI) in
the circumstellar envelope (CSE) of the carbon 
star IRC+10216. Our observations have uncovered a low surface brightness 
\HI\ shell of diameter $\sim1300''$ ($\sim$0.8~pc),
centered on IRC+10216. The \HI\ shell has an angular extent comparable to the far
ultraviolet-emitting astrosphere of IRC+10216 previously detected 
with the {\it GALEX} satellite, and its kinematics are consistent with
circumstellar matter that has been decelerated by the local interstellar
medium. 
The shell appears to completely surround the star, but the highest \HI\ column densities are
measured along the leading edge of the shell, near the location of a
previously identified bow shock.  We estimate a  total  mass of atomic hydrogen
associated with IRC+10216 CSE of  $M_{\rm HI}\sim
3\times10^{-3}~M_{\odot}$. 
This is only a small fraction of the
expected total mass of the CSE ($<$1\%) and is  consistent with the
bulk of the stellar wind originating in
molecular rather than atomic form, as expected for a cool star with an
effective temperature $T_{\rm eff}\lsim$2200~K. 
\HI\ mapping of a $2^{\circ}\times2^{\circ}$ region surrounding
IRC+10216 has also allowed us to characterize the line-of-sight
interstellar emission in the region and has uncovered a link between 
diffuse FUV emission southwest
of IRC+10216  and the Local Leo Cold Cloud. 
\end{abstract}


\section{Introduction}\protect\label{introduction}
At a distance of $\sim$130~pc, IRC+10216 (CW~Leonis) is the nearest 
example of a carbon-rich asymptotic giant branch (AGB) 
star. 
IRC+10216 is believed to be nearing the end of its evolution on the AGB and
to be close to transitioning into a protoplanetary nebula (e.g.,
Skinner et al. 1998; Osterbart et al. 2000). 
Its proximity and
advanced evolutionary state make
IRC+10216 an object of considerable interest for understanding
the late stages of evolution for intermediate mass stars. Some fundamental
properties of IRC+10216 are summarized in Table~1. 

As expected for a star approaching the end of its AGB lifetime,
IRC+10216 is undergoing mass loss at a high rate 
($\sim2\times10^{-5}~M_{\odot}$ yr$^{-1}$) and is surrounded by an
extensive circumstellar envelope (CSE). This CSE has  been widely
studied using a variety of observational tracers, including
dust-scattered
optical light (Mauron \& Huggins 1999); far infrared emission from
dust (e.g., Young et al. 1993; 
Ladjal et al. 2010; Decin et al. 2011; 
Groenewegen et al. 2012); far ultraviolet (FUV) continuum 
(Sahai \& Chronopoulos 2010); 
and spectral line emission from CO
(e.g., Knapp et al. 1998; Fong et al. 2003; Cernicharo et al. 2014) and a host of
other molecules (e.g., Olofsson et al. 1982; Cernicharo et al. 2000; 
Patel et al. 2009, 2011; De~Beck et al. 2012). 
Such studies have established
that the CSE of IRC+10216  is chemically complex and exhibits non-spherical 
structure on scales ranging
from milliarcseconds (e.g., Tuthill et al. 2000; 
Weigelt et al. 2002; Le\~ao et al. 2006) to tens
of arcminutes
(Sahai \& Chronopoulos 2010; Ladjal et al. 2010). IRC+10216 has a
moderately high space velocity (Table~1), and at large distances
from the star ($>$0.3~pc), 
there are also visible signatures of interaction between the 
outermost CSE and the interstellar medium (ISM), including
a bow shock ahead
of the star, and a wide-angle ``vortical tail'' trailing its direction
of space motion (Sahai \& Chronopoulos 2010; Ladjal et al. 2010). 

Despite the vast literature on the chemistry and morphology of the
IRC+10216 CSE, comparatively little is known about the predominant
component of the CSE by mass ($\sim$70\%)---the hydrogen gas. Outstanding
questions include
the total mass of the circumstellar hydrogen,
the fraction that is in atomic versus molecular form as a function of
distance from the star, and the relationship (both spatial and
kinematic) between the hydrogen gas
and the dust and other molecules. 
Such information is important for deciphering
the complex mass loss histories of stars like IRC+10216, including the 
amount of mass that they  return to the ISM while on the AGB, and the
details of their transition from AGB stars to planetary nebulae.

Glassgold \& Huggins (1983; hereafter GH83) showed that for a given mass loss rate, 
the fraction of an AGB star wind that is 
comprised of molecular hydrogen
versus atomic hydrogen is expected to depend largely
on the stellar effective temperature, $T_{\rm eff}$,
with the transition from  predominantly atomic to predominantly 
molecular winds predicted to occur for
$T_{\rm eff}\lsim 2500$~K. 
This is largely independent of whether the
star has a carbon-rich or oxygen-rich atmosphere.

Effective temperature determinations for IRC+10216 are notoriously uncertain
owing to the high opacity of the star's dusty
envelope, and values in the literature show considerable scatter. 
However, with the exception of Men'shchikov et
al. (2001), who proposed an effective temperature in the range
2500-3000~K based on a combination of radiative transfer and 
stellar evolutionary modelling, 
nearly all other published determinations lie within the
temperature range 
where molecular hydrogen is expected to comprise the bulk of the wind
and envelope: e.g.,
2230~K (Cohen 1979); 1800~K (Phillips et al. 1982); 
2330$\pm$350~K (Ridgway \& Keady 1988); 2200$\pm$150~K (Ivezi\'c \&
Elitzur 1996); 1915-2105~K (Bergeat et al. 2001). 

Observations to date have supported the suggestion that the bulk of the
IRC+10216 CSE is comprised of molecular rather than atomic gas (see
\S\ref{previous} for details), and indeed, radiation from H$_{2}$
molecules has 
been proposed as the
source of the extended FUV emission seen by {\it GALEX} (Sahai \& Chronopoulos
2010). However, even for stars as cool as
$\sim$2000~K, a thin shell of atomic hydrogen is predicted to form at
the edge of the CSE as a result of photodissociation of H$_{2}$ and
heavier molecules by the interstellar radiation field (Morris \& Jura
1983; GH83). For slightly warmer stars ($T_{\rm
  eff}\sim$2200~K), some small fraction of the hydrogen in the inner
CSE ($\lsim$1\%) is also expected be atomic  as a result of ``freeze-out'' of the
photospheric abundances   (GH83).
In the case of IRC+10216, both processes are predicted to produce
quantities of atomic hydrogen that are within the detection limits of
modern radio telescopes (GH83).
Furthermore, for stars like IRC+10216 that have appreciable space
velocities, additional neutral hydrogen may be
swept up from the ambient ISM
(Villaver et al. 2002, 2012). 
Motivated by these predictions, we have
undertaken a sensitive new
search for atomic hydrogen in the IRC+10216 CSE using 
\HI\ 21-cm line mapping 
observations obtained with the Robert C. Byrd Green Bank
Telescope (GBT) of the National Radio Astronomy
Observatory\footnote{The 
National Radio Astronomy
  Observatory is a facility of the National Science Foundation,
  operated under a cooperative agreement by Associated Universities,
  Inc.}. 

%%%%%TABLE 1%%%%%%%%%%%%%%%%%%%%%%%%%%%%%%%%%%%%%%%%%%%%%%%%%%%
\begin{table}
\begin{scriptsize}
\caption{Coordinates and Stellar Properties of IRC+10216}
\begin{tabular}{lcc}
\hline
Parameter & Value & Ref. \\
\hline

$\alpha$ (J2000.0) & 09 47 57.4 & 1\\

$\delta$ (J2000.0) & +13 16 43.5  & 1\\

$l$ & \ad{221}{45} & 1 \\

$b$ & \ad{+45}{06} & 1\\

Distance &  130~pc & 2\\
%Menten et al. 2012

Spectral Type & C9,5e & 3\\

Variability class & Mira & 4\\
%Ridgway \& Keady 1981

Pulsation period & 639$\pm$4 days & 5 \\
%Groenewegen et al. 2012\\

$T_{\rm eff}$$^{\rm a}$ & $\sim$2200~K &  see \S\ref{introduction}\\

Photospheric diameter (optical) & 3.8~AU & 2\\
%Menten et al. 2012\\

Luminosity & 8640$\pm430L_{\odot}$ & 2\\
%Menten et al. 2012\\

Initial Mass & 3-5~$M_{\odot}$ & 6\\
%Guelin et al. 1995

Current Mass & 0.7-0.9~$M_{\odot}$ & 7\\
%Ladjal et al. 2010

${\dot M}$
& 2$\times10^{-5}~M_{\odot}$ yr$^{-1}$ & 8\\
%Crosas \& Menten 1997

$V_{\rm outflow}$ & 14.6$\pm$0.3~\kms\ & 9\\
%Knapp et al. 1998

$V_{\rm LSR}$ & $-25.5\pm0.3$~\kms\ & 9\\
%Knapp et al. 1998 

Proper motion $(\mu_{\alpha}{\rm cos}\delta,\mu_{\delta})$ & 
(35$\pm$1,+12$\pm$1) mas yr$^{-1}$ & 2\\
%Menten et al. 2012

$V_{\rm space}$ & 42~\kms & 2\\
%Menten et al. 2012

P.A. of space motion & 70 degrees (E of N) &  2\\
%Menten et al. 2012

\hline
\end{tabular}

{Units of right ascension are hours, minutes, and
seconds, and units of declination are degrees, arcminutes, and
arcseconds. All quantities have been scaled to the distance adopted in
this paper.

$^{a}$ Based on a mean of various determinations in the literature.

(1) SIMBAD database; (2) Menten et al. 2012; (3) Cohen
  1979; (4) Ridgway \&
  Keady 1981; (5) Groenewegen et al. 2012; (6) Guelin et al. 1995; (7)
  Ladjal et al. 2010; (8) Crosas \& Menten 1997; (9) Knapp et
  al. 1998  }
\end{scriptsize}
\end{table}
%%%%%%%%%%%%%%%%%%%%%%%%%%%%%%%%%%%%%%%%%%%%%%%%%%%%%%%%%%%%%%

\section{Previous \HI\ 21-cm Line Observations of IRC+10216}\protect\label{previous}
Several previous authors have attempted to detect neutral atomic hydrogen
associated with
the CSE of IRC+10216. Here we briefly summarize these
efforts to provide
context for the current study. All stellar parameters quoted in this section 
have been scaled
to the distance adopted in this paper.

Zuckerman et al. (1980) were the first to attempt detection 
of \HI\ 21-cm line emission 
associated with mass loss from IRC+10216.  Using the Arecibo telescope (which has a 
FWHM beamwidth at 21-cm of \am{3}{2}), Zuckerman et al.
obtained a position-switched
(on$-$off) measurement with the ``on'' position centered on the star and
the ``off'' measurement comprising an average of four pointings
displaced by $\pm$\am{3}{5} in RA and $\pm$\am{3}{5} in DEC,
respectively. No circumstellar \HI\ emission was detected, and these authors
placed a (1$\sigma$) upper limit on the \HI\ mass within the 
Arecibo beam of $M_{\rm HI}<6.3\times10^{-4}~M_{\odot}$.

Using observations obtained with the Very Large Array (VLA) in its C
configuration, Bowers \&
Knapp (1987) also failed to detect any 
\HI\ emission associated with IRC+10216. These authors
placed a $3\sigma$ upper limit on the \HI\ mass within a synthesized
beam (50$''$ FWHM) centered on 
the stellar position of $M_{\rm HI}<5.5\times10^{-5}M_{\odot}$ and a $3\sigma$
upper limit on the total CSE mass 
of $M_{\rm HI}<1.0\times10^{-4}~M_{\odot}$ assuming a source
diameter of $6'$.

%%%%%%%%%%%%%%%%%%%%%%%%%%%%%%%%%%%%%%%%%%%%%%%%%%%%%%%%%%%%%%%%%%
\setcounter{figure}{0}
% Fig. 1
% 
\begin{figure}
\includegraphics[scale=0.4,angle=-90]{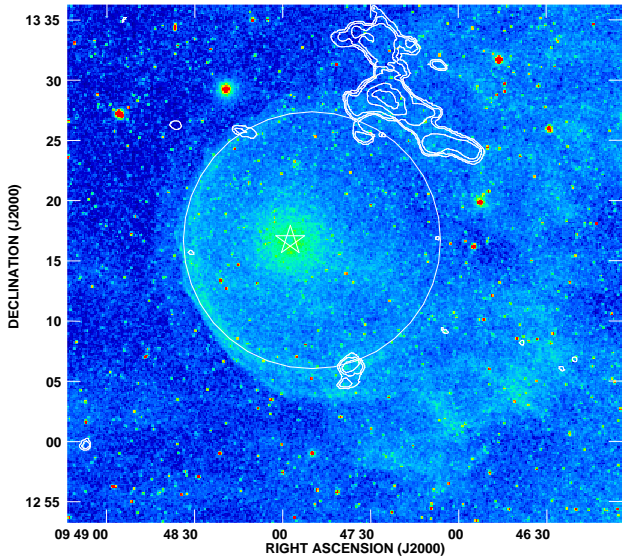}
\caption{{\it GALEX} FUV image of IRC+10216 with VLA \HI\ contours
 from MR07 overplotted. The contour levels are
 (6,12,18,38,48)$\times$1.2~Jy beam$^{-1}$ m s$^{-1}$. No correction
 for the VLA primary beam has been applied. The VLA
 synthesized beam was $\sim101''\times94''$. A white star symbol
 indicates the stellar position. The overplotted circle
 has a diameter of 1280$''$ and has
been displaced by 110$''$ west of the star (see
 \S\ref{windHI} for details). }
\label{fig:vlaoverlay}
\end{figure}

%%%%%%%%%%%%%%%%%%%%%%%%%%%%%%%%%%%%%%%%%%%%%%%%%%%%%%%%%%%%%%%%%%%

While these earlier results were somewhat discouraging, interest in
detecting circumstellar hydrogen around evolved stars was
reinvigorated roughly a decade later when
Le~Bertre \& G\'erard (2001) revisited IRC+10216 using the then newly upgraded
\nan\ Radio Telescope (NRT). The NRT 
has an elongated beam with a FWHM at 21~cm of $\sim4'$
east-west and $\sim22'$ north-south. Based on position-switched
measurements where the ``off'' spectra comprised of the
average of spectra
displaced by $\pm4'$, $\pm8'$, and
$\pm12'$ in RA from the stellar position,  respectively, 
Le~Bertre \& G\'erard found a surprising result:
an \HI\
{\em absorption} profile centered at the stellar systemic velocity,
with a velocity width comparable to twice the outflow speed of the
star.  They attributed this finding to 
cold circumstellar in the outer CSE, seen in absorption against the
cosmic microwave background radiation (but see below). 

Motivated by the results of Le~Bertre \& G\'erard (2001), 
Matthews \& Reid (2007; hereafter MR07) 
obtained new \HI\ imaging observations of
IRC+10216 with the VLA,  this time using its most compact (D)
configuration to improve sensitivity to extended emission. 
Compared with the earlier observations of Bowers \&
Knapp (1987), the data of MR07 also had $\sim$2.5 times lower RMS noise
(normalized to a 50$''$ synthesized beam). Consistent
with Bowers \& Knapp, MR07 detected no \HI\ 
emission directly toward the stellar position. 
However,  MR07 did find several
isolated clumps and arcs of emission displaced by $\sim$11$'$ to 18$'$
($\sim$0.4 to 0.7~pc in projected distance) 
from the star. The velocities of these features 
are consistent with 
molecular gas previously observed at smaller radii  in the CSE ($r\lsim 120''$),
leading MR07 to suggest that they could plausibly be circumstellar in origin. 
Assuming that the clumps and arcs detected with the VLA
between LSR velocities $-33$~\kms\ and $-17$~\kms\ 
were associated with the CSE, MR07 derived a circumstellar \HI\ mass of
$\sim 2.2\times10^{-3}~M_{\odot}$. 
% MR07 used distance of 135 pc
The results of the VLA imaging also led MR07
to suggest an alternative interpretation for the apparent absorption
spectrum seen by Le~Bertre \& G\'erard (2001)---namely that the
off-source spectra used for the position-switched NRT measurements 
may have been contaminated by \HI\ {\em emission}, thus 
mimicking an absorption signature in on$-$off
difference spectra. 

A further twist came three years later, with the discovery of
a FUV-emitting astrosphere surrounding IRC+10216 (Sahai \&
Chronopoulos 2010). Figure~\ref{fig:vlaoverlay} 
shows a version of the FUV image with the VLA \HI\ contours of MR07 
overplotted. As pointed out by Matthews et al. (2011), 
this overlay is highly suggestive of a relationship  
between the \HI\
clumps and arcs detected with the VLA and the circumstellar
structures traced in FUV light by {\it GALEX}. 
However, the bulk of the \HI\ emission
detected with the VLA lies near, or just outside the
$\sim30'$ FWHM VLA primary beam, where sensitivity drops
steeply, and a
chance confusion with an interstellar cloud along the line-of-sight
cannot be strictly excluded. 
At the same time, an interferometer like the VLA has rather poor
sensitivity to diffuse, low surface brightness emission that is
extended over scales of several arcminutes or more.  This 
raises the possibility that  
there may be additional atomic hydrogen present in the IRC+10216
CSE that was not detected with the VLA, but which could be detectable
through sensitive observations with a single-dish telescope.

For these reasons, we have embarked on a new investigation of IRC+10216 in
the \HI\ 21-cm line using the GBT.  A combination of several attributes of 
the GBT make it particularly well-suited
for such a study. Its 100-m aperture provides exceptional
sensitivity, and its $\sim9'$ FWHM beamwidth
at 21~cm is comparable to the largest angular scales detectable in the
VLA D configuration data of MR07. Furthermore, the extremely low
sidelobe levels of the GBT main beam, coupled with 
the availability of new software 
to make robust stray
radiation corrections (Boothroyd et al. 2011),  insure that low-level extended 
emission can be reliably detected and
characterized. Finally, unlike the NRT, which is a transit instrument,
the GBT can perform 
on-the-fly mapping, allowing us to efficiently map the entire extended
astrosphere and tail of
IRC+10216 and also characterize the ISM across an
extended region surrounding the star.

Shortly after our new GBT  data were obtained, Menten et al. (2012)
performed an investigation of the region around IRC+10216 using
\HI\ data from the Arecibo GALFA-\HI\ survey (Peek et al. 2011a) 
and reported finding no sign of \HI\ emission associated with the CSE.  
However, the GBT data have a surface brightness sensitivity
$\sim$7 times higher than
the GALFA-\HI\ data, and as described below (\S\ref{results}), 
this improved sensitivity
coupled with a more extensive analysis, has yielded some new information, 
including the detection of neutral hydrogen that appears to
be associated with the IRC+10216 CSE.

\section{Observations and Data Reduction}\protect\label{observations}
\HI\ 21-cm spectral line mapping of a 
$2^{\circ}\times2^{\circ}$ region surrounding IRC+10216 was obtained using
the GBT during a series of seven observing sessions on 2011 
November 16, 17, 20, 22, and 29 and 2011 December
6 and 7. 
Pointing and focus were checked at the start of each session using
observations of a suitable continuum source. For the spectral line
observations, 
the GBT spectrometer was employed with a 12.5~MHz bandwidth and 9-level
sampling. In the raw data,
there were 16,384 spectral channels with a 0.7629~kHz (0.16~\kms)
channel spacing.
In-band frequency switching was used with cycles of 0.8~Hz, alternating
between frequency shifts of 0 and $-2.5$~MHz from the
centre frequency of 1420.52~MHz ($V_{\rm LSR}=-25.5$~\kms). 
This resulted in a usable LSR velocity
range from $\sim-500$~\kms\ to $+500$~\kms. Data were recorded in dual linear
polarizations. 

System temperatures over the course of our observing runs ranged from 11~K to
24~K, and mean values for each of the seven observing sessions 
ranged from 16.3~K to 18.6~K.
Absolute calibration of the brightness temperature scale was determined from
injection of a noise diode signal at a rate of 0.4~Hz  and
was checked for consistency during each session with
observations of the line calibrator S6
(Williams 1973).

To maximize observing efficiency, we employed the on-the-fly (OTF)
mapping technique (Mangum et al. 2007). Given the spatial extension 
of the IRC+10216
astrosphere and trailing wake previously seen in the FUV 
(Figure~\ref{fig:vlaoverlay}), 
we adopted a map centre  \ad{0}{25} west of the stellar position 
($\alpha_{\rm J2000}=09^{\rm h} 47^{\rm m} 00^{\rm s}$,
$\delta_{\rm J2000}=13^{\circ} 16'$\as{43}{56}).  

At the frequency of our observations, the FWHM of
the GBT beam is $\theta\sim$\am{8}{7}. We therefore used a
spatial sampling of \am{1}{62} (approximately two times higher than the
Nyquist value of $\theta/2.4$) to avoid the beam degradation and decrease in the 
signal-to-noise ratio that occur with coarser sampling 
(see Magnum et al. 2007). 
We scanned in right ascension, and obtained several
complete passes over the entire $2^{\circ}\times2^{\circ}$ 
target region. Because observing time for our
program was
allocated dynamically, the size of the region mapped, the scan rate,
and the dump time
were adjusted  from session to session to insure that we achieved a
suitable balance between our two objectives: (1) obtaining
a high
sensitivity map across the IRC+10216 astrosphere; and (2) achieving 
sufficient spatial
coverage to characterize the larger scale environment of the star.  
Scan rates ranged
from 12 arcsec s$^{-1}$ to 97 arcsec s$^{-1}$ and 
dump rates between 1 and 4 seconds. A consequence is that
our resulting map has a lower RMS noise level in the
inner \ad{1}{0}$\times$\ad{1}{0}  compared with the outer
\ad{0}{5} border (see below). In total, 42.6 hours of data were used
to construct our final map.

Initial data reduction steps were performed using 
GBTIDL\footnote{http://gbtidl.nrao.edu/}. 
The total power for each scan was computed using
two reference spectra (during which the noise diode
was fired) and two signal spectra. After combining these, the calibrated 
data were folded to average the two
parts of the in-band frequency-switched spectrum. A handful of
spectral channels that showed persistent radio frequency interference
were flagged, and the data were then smoothed in frequency with a
boxcar function (of kernel width 5) and decimated, resulting in spectra
with a channel spacing of $\sim$0.8~\kms. 

At this stage, the data were corrected for stray radiation using 
the approach developed by Boothroyd et al. (2011). These corrections are
expected to reduce systematic errors in 21-cm line measurements with
the GBT by as
much as an order of magnitude, and are thus particularly important 
for studies like ours 
that aim to detect and characterize weak, extended emission,

Following the corrections for stray radiation, the data were imported
back into GBTIDL. 
The flux density scale was converted from units of
main beam brightness temperature  to
Janskys assuming a gain factor of 0.463 Jy K$^{-1}$, and
a third order polynomial baseline was fitted and subtracted from
each spectrum. The regions of the spectrum used for the baseline fits
correspond to 
$-275 \lsim V_{\rm LSR}\lsim -120$~\kms\ and $133 \lsim V_{\rm LSR}
\lsim 287$~\kms. 

The baseline-subtracted spectra were converted to standard SDFITS format using
the idlToSdfits program developed by G. Langston and subsequently 
imported into the
Astronomical Imaging and Processing Software package 
(AIPS; Greisen 2003) for further
processing and analysis. Within AIPS, the data from the seven different
observing sessions were concatenated and then convolved and sampled
onto a regular grid using the task {\sc SDGRD}, resulting in a
three-dimensional spectral line data cube. For the gridding, a
Bessel$*$Gaussian convolution function was used with
cell size of 100$''$. Other gridding parameters were optimized as per
the recommendations of Mangum et al. (2007). The RMS
noise in the resulting data cube is $\sim$5.5~mJy beam$^{-1}$ per
0.8~\kms\ spectral channel
within the central
\ad{1}{0}$\times$\ad{1}{0} of the maps
and $\sim$20~mJy beam$^{-1}$ along the outer \ad{0}{5} border.

%%%%%%%%%%%%%%%%%%%%%%%%%%%%%%%%%%%%%%%%%%%%%%%%%%%%%%%%%%%%%%%%%%
% Fig. 2
% 
\begin{figure}
\vspace{-1.5cm}
\includegraphics[scale=0.37, angle=90]{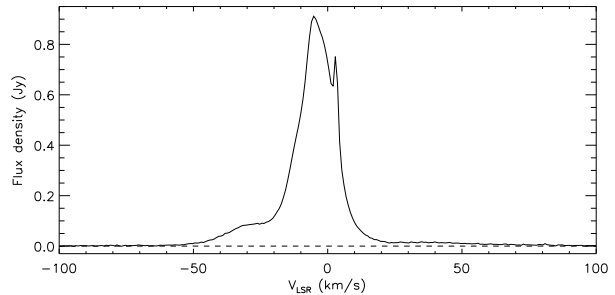}
\caption{Frequency-switched 
GBT \HI\ spectrum toward the position of IRC+10216. The emission was
integrated over a 9-pixel ($300''\times300''$) region centered on the
stellar position. }
\label{fig:onstar}
\vspace{-0.5cm}
\end{figure}
%%%%%%%%%%%%%%%%%%%%%%%%%%%%%%%%%%%%%%%%%%%%%%%%%%%%%%%%%%%%%%%%%%

\section{Results}\protect\label{results}
Despite the moderately high Galactic latitude of IRC+10216, \HI\ line
profiles toward its direction are complex and exhibit signatures of
multiple interstellar emission components along the line-of-sight
(Figure~\ref{fig:onstar}; see also Hartmann \& Burton 1997). 
These interstellar emission
components must be disentangled to allow identification 
of the much weaker circumstellar emission. 

As an initial step toward characterizing the line-of-sight emission, we 
performed Gaussian decompositions of 
frequency-switched spectra 
near the position of IRC+10216. 
This analysis has yielded insights into the local
interstellar environment of IRC+10216 as well as the other
interstellar components toward this direction, including 
high- and intermediate-velocity gas and an emission component
associated with the Local Leo Cold Cloud. Details are provided in
Appendix. However,  
these spectral decompositions did not allow us
to unambiguously identify
any \HI\ emission associated with the circumstellar environment of IRC+10216.
For this purpose, it was necessary to take further advantage of the spatial
information provided by our spectral mapping, including examination of \HI\
channel maps and moment maps and constructing 
difference spectra using observations toward the CSE of IRC+10216 and
at neighboring
reference positions. The results of this analysis are described in the
sections that follow.

\setcounter{figure}{2}
%%%%%%%%%%%%%%%%%%%%%%%%%%%%%%%%%%%%%%%%%%%%%%%%%%%%%%%%%%%%%%%%%%
% Fig. 3a
% 
\begin{figure*}
\vspace{0.5cm}
\includegraphics[scale=0.89]{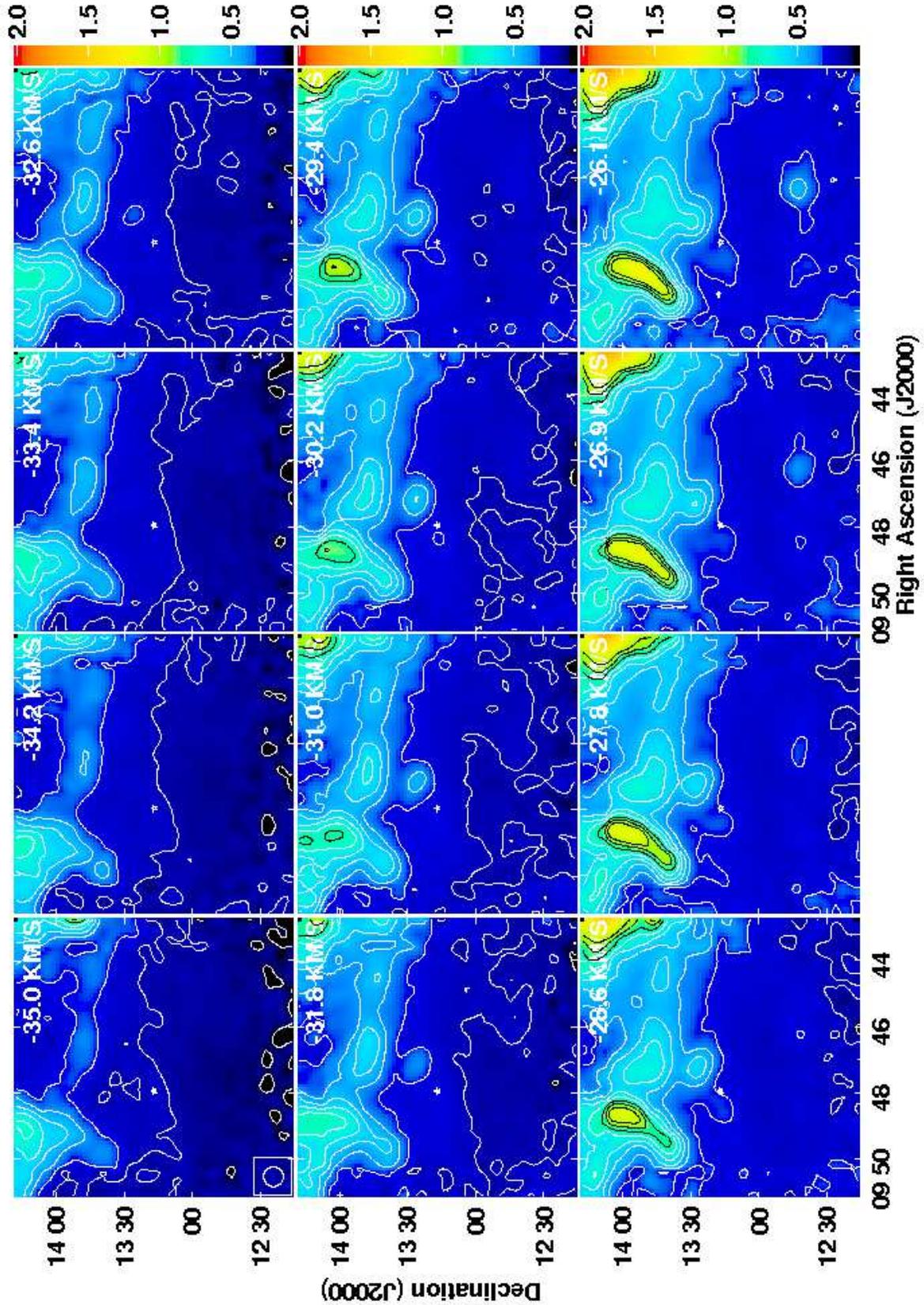}
\caption{GBT \HI\ channel maps at LSR 
velocities close to the stellar systemic
 velocity of IRC+10216 ($V_{\rm LSR, \star}=-25.5$~\kms.) 
Contour levels are (1,2,...10)$\times$0.1 Jy beam$^{-1}$. The intensity scale
has units of Jy beam$^{-1}$. A star symbol indicates the stellar
 position. 
 } 
\label{fig:cmaps}
\end{figure*}

% Fig. 3b
% 
\begin{figure*}
\vspace{0.5cm}
\includegraphics[scale=0.89]{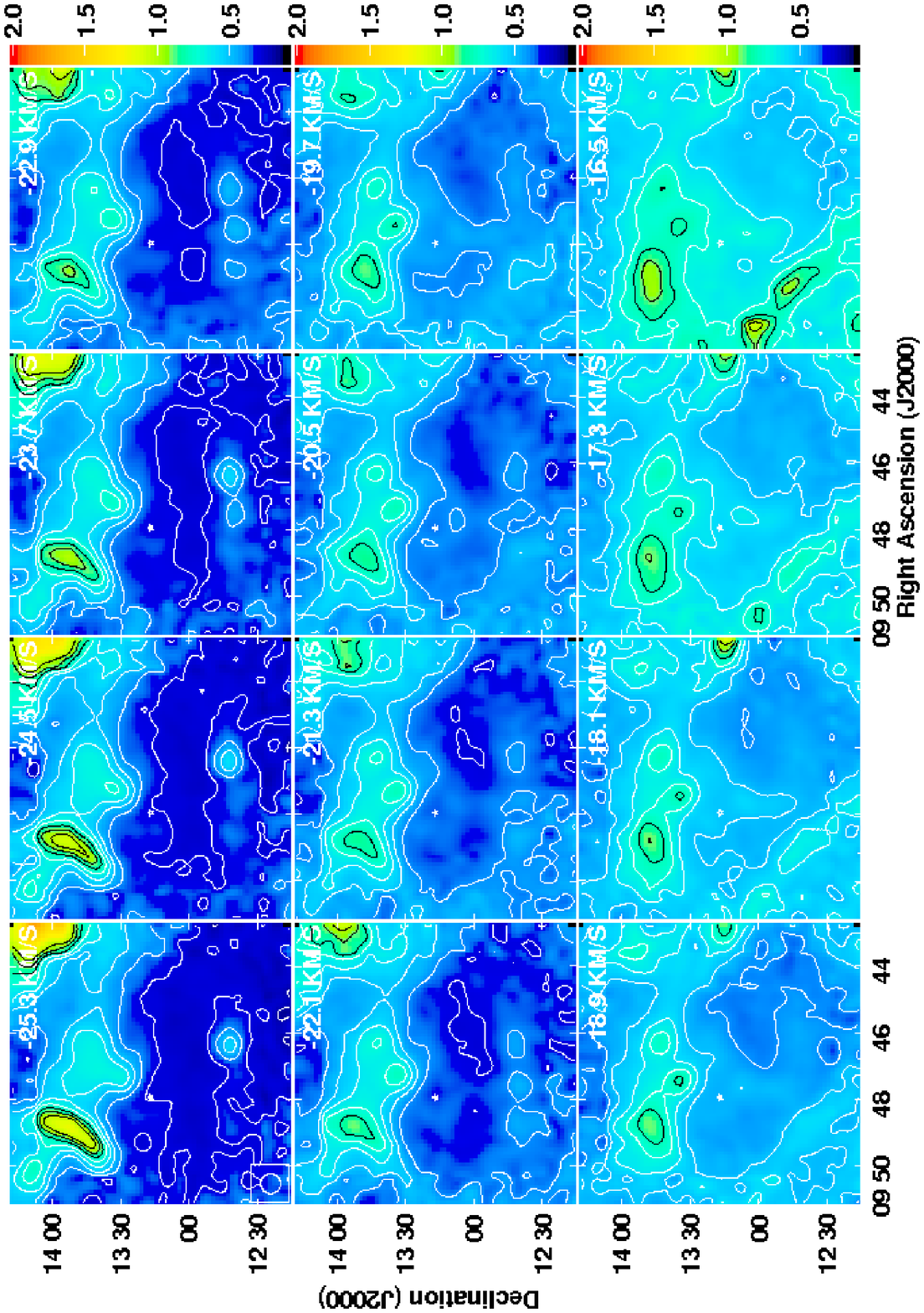}
\contcaption{}
\end{figure*}

%%%%%%%%%%%%%%%%%%%%%%%%%%%%%%%%%%%%%%%%%%%%%%%%%%%%%%%%%%%%%%%%%%

\subsection{Identification of \HI\ Emission Associated with the
  IRC+10216 Astrosphere}\protect\label{windHI}
Figure~\ref{fig:cmaps} presents a series of \HI\ channel maps
at velocities bracketing the systemic velocity of IRC+10216. We see that 
the bulk of the
emission at these velocities lies in a spatially 
extended complex of clouds 
north of IRC+10216 that is almost certainly 
unrelated to mass loss from IRC+10216.  Also present at
these velocities is a set of three compact clouds south of 
IRC+10216 that appear to lie
along a single linear filament
near declination 12$^{\circ}$ 41$'$ and between right ascension
09$^{\rm h}$ 46$^{\rm m}$
and 09$^{\rm h}$ 49$^{\rm m}$. These clouds too are likely 
unrelated to the star.\footnote{These clouds
appear qualitatively
similar to the types of condensations expected in post-shock
gas following the collision of a high-velocity gas cloud with the
Galactic disc (Tenorio-Tagle et al. 1986).
These features may therefore hold clues on the origin
of the Local Leo Cold Cloud and the high- and intermediate-velocity gas components
that all occur toward this region of the sky (see Stark et al. 1994 
and the Appendix).}

Although line-of-sight emission dominates the channel maps shown in 
Figure~\ref{fig:cmaps},  inspection of the region near
IRC+10216 reveals something interesting:
evidence of a faint, shell-like structure surrounding
the stellar position. A nearly complete shell is visible 
in several consecutive velocity channels redward
of the stellar velocity, most notably in the channels centered at
$V_{\rm LSR}=-22.1$, $-21.3$, $-20.5$, and $-19.7$~\kms. 
The approximate diameter of the
shell is $\sim1280''$, or $\sim$0.8~pc at the distance of IRC+10216. 
An arc of emission
consistent with the shell's location is also seen northeast of the star in
velocity channels spanning 
$-30.2\le V_{\rm LSR}\le -23.7$~\kms. Redward of
$V_{\rm LSR}=-18.9$~\kms, 
background confusion begins to dominate and signatures of the shell are no longer
readily apparent. 

\begin{figure}
\includegraphics[scale=0.42]{LMatthews_fig4a.ps}
\includegraphics[scale=0.42]{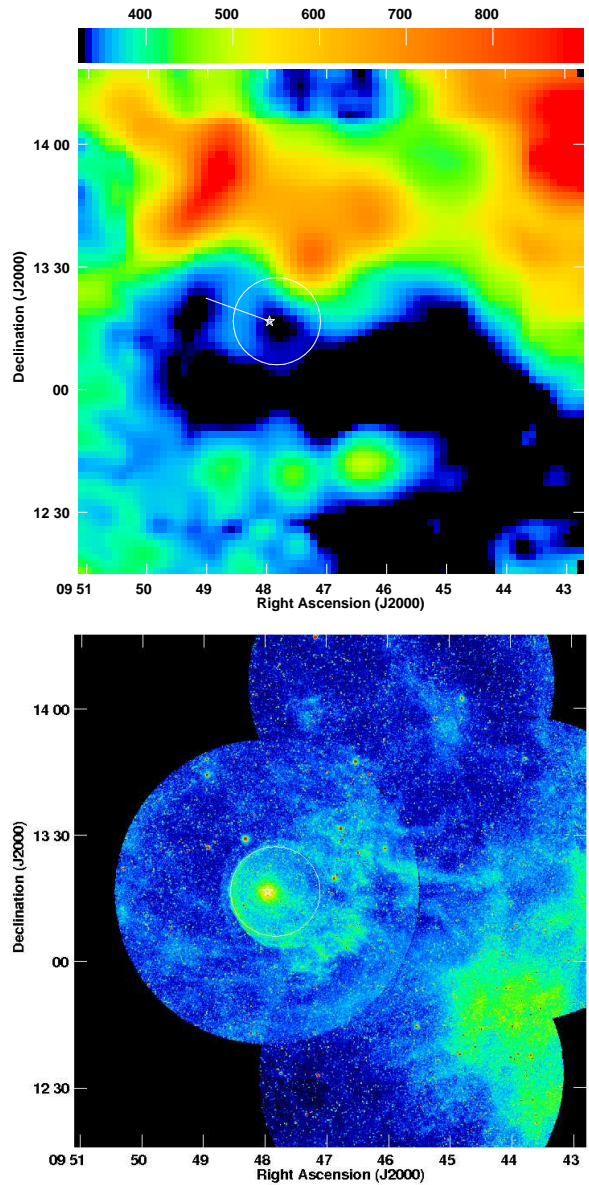}
\caption{{\it Top:} \HI\ intensity map obtained by averaging 
the emission
over the velocity range $-24.5 \le V_{\rm LSR} \le
-21.3$~\kms.  The
image is displayed using
square root  intensity scaling in units of 
mJy beam$^{-1}$. A star symbol denotes the position of
IRC+10216, and the solid white line indicates its
direction of space motion.
{\it Bottom:} A {\it GALEX} FUV mosaic of the same field shown in the
top panel. Identical
circles with diameters of 1280$''$ are overlaid on both panels to 
highlight the location of 
shell-like structures visible at the respective
wavelengths. The centres of the circles have
been displaced by 110$''$ west of the stellar position. }
\label{fig:rings}
\end{figure}

To further highlight the putative shell, in 
Figure~\ref{fig:rings} (top panel) we show an image produced by taking
a mean of the
emission over the velocity range $-24.5 \le V_{\rm LSR} \le
-21.3$~\kms. In the bottom panel, we show on the same scale a {\it GALEX} FUV
image of the region. Despite confusion with line-of-sight emission
along the northern and western part of the shell, we see that it matches
closely in both size and position with the FUV-emitting astropause
surrounding IRC+10216. The geometrical centre of both the \HI\ and FUV
shells are also displaced by a comparable amount ($\sim110''$) west of
the stellar position (see also Figure~\ref{fig:vlaoverlay}). The implications
of this are discussed in \S\ref{discussion}.
 
To estimate the mean, beam-averaged
\HI\ column density of the shell, we have computed a sum of channel
images from $-24.5 \le V_{\rm LSR} \le
-21.3$~\kms\ and integrated the resulting map over a semi-circular
aperture with radius $r=900''$, centered 110$''$ west of the stellar
position, and extending between position angles of $+70^{\circ}$ to
$+250^{\circ}$. The remaining position angles were
excluded because of obvious contamination from a cloud complex
north of the star. We find
$N_{\rm HI}< 2.1\times10^{19}$~cm$^{-2}$. This value is formally 
an upper limit over the range of velocities 
used to construct our total intensity map because the shell is 
superposed on a pedestal
of background emission. However, the total column density associated
with the shell may be higher if additional circumstellar material is
present at higher
LSR velocities where it is obscured by line-of-sight confusion.

A region of enhanced surface brightness is visible along the
northeastern edge of the shell in Figure~\ref{fig:rings}, upstream
of the star's direction of space motion (along a position
angle of 70$^{\circ}$; Table~1). 
The enhanced \HI\ column density at this location is consistent
with a build-up of atomic hydrogen near the bow
shock region previously identified in FUV and FIR images
(Sahai \& Chronopoulos  2010; Ladjal et al. 2010). The observed region
of enhanced \HI\ column density
also spans a similar range of position angles as the region
where Sahai \& Chronopoulos identified
a sharp falloff in the FUV emission intensity, which they attributed to 
gas pile-up outside
the astropause. Our new observations imply that not all of the gas in
this vicinity is ionized, despite sufficiently high predicted gas
temperatures. Assuming a strong shock approximation and that the
bow shock is sweeping up a mostly atomic material from the surrounding
ISM, then
$T\sim\left(\frac{3}{16}\right)\mu m_{\rm H}V^{2}_{\rm space}/k\approx 52,000$~K, where
$\mu$=1.3 is the mean molecular weight of an atomic gas comprising
90\% H and 10\% He, $m_{\rm H}$ is the mass of the H atom,
$k$ is the Boltzmann constant, and $V_{\rm
  space}\approx42$~\kms\ is the space velocity of the star.

\begin{figure}
\includegraphics[scale=0.48]{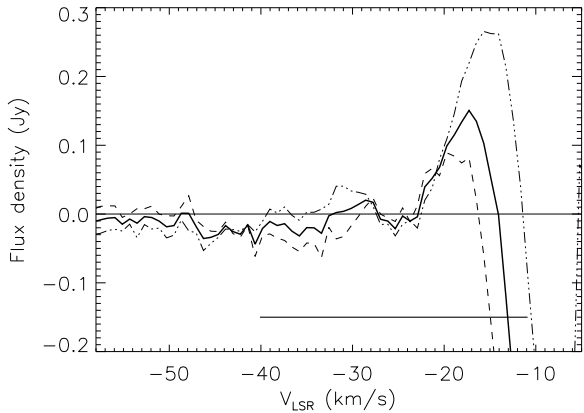}
\caption{Position-switched
spectrum of the \HI\ shell. An on-source spectrum was constructed by 
integrating within a semi-circular
aperture of radius $900''$, centered on IRC+10216 and differenced with
reference spectra extracted east of the star (dashed line) and west of
the star (dot-dash
line), respectively. See
text for additional details.  The thick line shows the difference spectrum
obtained using a mean of the east and west reference positions. 
The horizontal bar indicates the expected range of gas
velocities for the IRC+10216 CSE based on the outflow velocity
measured from CO.
 }
\label{fig:ringspec}
\end{figure}

While the \HI\ shell in  Figure~\ref{fig:rings} is visible in channel
images created from frequency-switched data, as noted above, it is superposed on a
background of line-of-sight emission that is widespread
in this region (see also \S\ref{IVG}). 
To estimate the mass of the shell, we therefore
have derived one-dimensional position-switched spectra by integrating over the
entire shell in each individual channel map 
and differencing the result with neighboring reference
spectra.  
For the on-source spectrum, we integrated the emission over the 
same semi-circular aperture as used for the column density
determination described
above. We then constructed the reference (``off'') spectra by integrating in
$600''\times1700''$ rectangular regions centered at positions offset
$(+1200'',-100'')$ and $(-1500'',-100'')$ from the stellar
position, respectively.  These positions 
are offset one beamwidth east and west, respectively, from the 
outer boundaries of the \HI\ shell, while the
north-south extent of the reference apertures matches that of the
semi-circular on-source aperture.
To account for the small difference in integration 
area, a correction factor was applied to each 
reference spectrum before differencing with 
the on-source spectrum. The resulting spectra are shown in
Figure~\ref{fig:ringspec}.

From the spectrum in Figure~\ref{fig:ringspec} we computed the velocity-integrated 
\HI\ flux density by taking the area underneath the resulting
difference spectrum between LSR velocities of $-29.4$~\kms\
and $-18.1$~\kms\ (the velocity range over which emission at the
location of the shell is visible in the individual channel maps). We find 
$\int S_{\rm HI} {\rm d}v\approx$0.40$\pm$0.04~Jy~\kms. The quoted
error bar reflects the difference in the values obtained using the
east and the west off-source spectra. To obtain an 
estimate of the total \HI\ mass of the shell, we assume the shell is
symmetric about the position angle of space motion (PA=$70^{\circ}$) and
multiple our measured
value by a factor of two to account for the fact that the 
flux density was integrated over
only half the shell. We also assume that no emission is present at
higher LSR velocities where contamination begins to dominate. These assumptions
may lead to an uncertainty in our derived \HI\ mass by as much as a factor of
two. Thus our best estimate for the mass of the \HI\ shell is 
$M_{\rm HI}\sim(3.2\pm1.6)\times10^{-3}~M_{\odot}$. 

\begin{figure}
\includegraphics[scale=0.48]{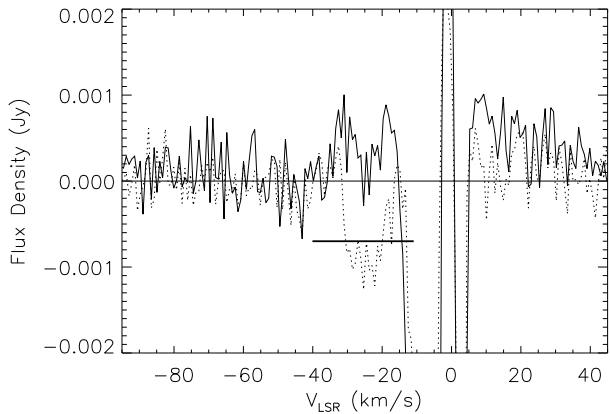}
\caption{Position-switched
spectra along the line-of-sight toward IRC+10216. 
An on-source spectrum was
 extracted over a singe $100''$ pixel at the stellar
 position and differenced with reference spectra
extracted $1100''$ east (outside the \HI\ shell; 
solid line) and $600''$ east (along the \HI\ shell; dotted line). In
the latter spectrum, a negative feature is seen corresponding
to the expected velocity range of the \HI\ shell. The horizontal
bar indicates the expected range of gas
velocities for the IRC+10216 CSE based on the outflow velocity
measured from CO.
 }
\label{fig:positionswitched}
\end{figure}

The \HI\ shell seen in Figure~\ref{fig:rings}a appears to have a
central depression, and consistent with previous workers (\S\ref{previous}), we find no
compelling evidence of \HI\ emission directly along the line-of-sight
to IRC+10216.
In  Figure~\ref{fig:positionswitched} we show difference
spectra where the ``on'' spectrum was integrated over a
single 100$''$ pixel centered on IRC+10216, and the
reference spectra were extracted $1100''$ east (just outside the \HI\
shell) and $600''$ east (along the \HI\ shell)
respectively.
In the former case, there is no significant emission at the
stellar systemic velocity, although there are hints of emission peaks
on either side of $V_{\rm sys}$, within the velocity range expected
for circumstellar gas (as
indicated by the horizontal bar).  Although this 
line profile is consistent with the type of 
``double-horned'' profile expected for optically thin emission that
fills the beam (e.g., Zuckerman 1987), we do not take this
detection to be significant given that the blue wing of the line overlaps in
velocity 
with the intermediate velocity gas that
is widespread throughout the region (\S\ref{IVG}), while the red edge
of the profile ($V_{\rm  LSR}\gsim -18$~\kms)
is strongly contaminated by line-of-sight emission. 
Further, in the difference spectrum constructed with a reference
position along the shell, the blue and red peaks disappear, and 
we see instead only an negative spectral feature whose velocity
spread is consistent with the velocity extent of the shell at this
position in the \HI\ channel maps (Figure~\ref{fig:cmaps}).

Integrating  the first spectrum in  Figure~\ref{fig:positionswitched} 
across the velocity
range of the CSE (but excluding velocities $V_{\rm
  LSR}\ge -18$~\kms) we derive an integrated \HI\ flux density of
$\sim0.0078$Jy~\kms, translating to an \HI\ mass of
$M_{\rm HI}\le 3\times10^{-5}~M_{\odot}$. This is consistent with
previous upper limits toward the stellar position reported by other
authors (see \S\ref{previous}). The
simplest interpretation of these results is that the inner portions of
the IRC+10216 are predominantly molecular, with at most a small atomic
fraction. If we assume that the standard mass continuity equation
holds for the inner wind region (e.g., Lamers
\& Cassinelli 1999) and adopt $r=9.7\times10^{16}$~cm
(50$''$), then $M_{\rm tot}\approx \frac{1}{3}{\dot M}r/V_{\rm out} =
0.014M_{\odot}$ using the values of ${\dot M}$
and $V_{\rm out}$ from Table~1. Thus $M_{\rm HI}/M_{\rm tot}\le$0.2\%.
According to GH83, an atomic wind fraction of this magnitude can be
explained by photospheric freeze-out if the stellar effective
temperature is $\sim$2200~K, whereas the atomic fraction should become
negligible if $T_{\rm eff}\sim$2000~K (see also \S\ref{discussion}).

%%%%%%%%%%%%%%%%%%%%%%%%%%%%%%%%%%%%%%%%%%%%%%%%%%%%%%%%%%%%%%%%%%
% Fig. 4
% 
%%%%%%%%%%%%%%%%%%%%%%%%%%%%%%%%%%%%%%%%%%%%%%%%%%%%%%%%%%%%%%%%%%

%%%%%%%%%%%%%%%%%%%%%%%%%%%%%%%%%%%%%%%%%%%%%%%%%%%%%%%%%%%%%%%%%%
% Fig. 5
% 
%%%%%%%%%%%%%%%%%%%%%%%%%%%%%%%%%%%%%%%%%%%%%%%%%%%%%%%%%%%%%%%%%%

%%%%%%%%%%%%%%%%%%%%%%%%%%%%%%%%%%%%%%%%%%%%%%%%%%%%%%%%%%%%%%%%%%
% Fig. 6
% 
%%%%%%%%%%%%%%%%%%%%%%%%%%%%%%%%%%%%%%%%%%%%%%%%%%%%%%%%%%%%%%%%%%

\subsection{\HI\ Velocity Dispersion Map}
\begin{figure}
\includegraphics[scale=0.4]{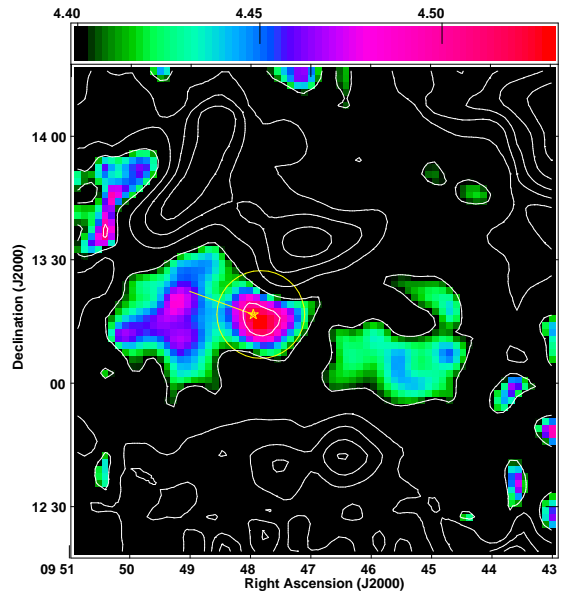}
\caption{\HI\ velocity dispersion map of the region around IRC+10216
  (shown as both contours and color-scale),
constructed by taking the intensity-weighted second moment of data spanning
 velocities $-32.5 \le V_{\rm LSR} \le -18.0$~\kms. The maximum velocity
 dispersion in the region is coincident with the location of the star. The overplotted
 circle is as in Figure~\ref{fig:rings}, and the yellow line indicates
 the direction of space motion.  Contour levels span 3.8 to
 4.5~\kms\ in increments of 0.1~\kms. With the adopted intensity
 stretch, regions on the map with values $\le4.4$~\kms\ appear black. Because of the
 imposed velocity cutoffs used to construct the map, 
absolute values of the velocity dispersion are not physically meaningful. 
 }
\label{fig:veldisp}
\end{figure}

\protect\label{dispersion}
In Figure~\ref{fig:veldisp} we present
an \HI\ velocity dispersion (second moment) map derived using data over
the velocity range $V_{\star,\rm LSR}\pm7.3$~\kms. This corresponds to the
inner 50\% of the range of gas velocities detected in molecular lines and
is approximately the velocity range over which the \HI\ shell is seen
in our data. (Note however that this velocity range is larger than used to
construct the \HI\ intensity map in Figure~\ref{fig:rings}a). 
Because of the imposed velocity cutoffs used to construct
this velocity dispersion  map, the absolute values of the gas 
dispersion are not physically
meaningful. Nonetheless, this map
offers a tool to disentangle emission related to the
circumstellar environment of IRC+10216 from other line-of-sight
emission. 
In particular, we find a
coincidence between the peak 
velocity dispersion in the region ($\sigma_{V}=4.5$~\kms) with the position of
IRC+10216. This location
has a small ($\sim$0.2~\kms) but significant enhancement in
velocity dispersion relative to the mean of its
surroundings. 
Additional patches of slightly enhanced velocity dispersion are also
seen both
upstream and downstream of IRC+10216, along the trajectory of space
motion of the star. It is unclear if these are related to the motion
of the star through the ISM, although in general, a region of enhanced turbulence
may be expected within a foreshock zone lying just ahead of the bow
shock (e.g., Blandford \& Eichler 1987).

%%%%%%%%%%%%%%%%%%%%%%%%%%%%%%%%%%%%%%%%%%%%%%%%%%%%%%%%%%%%%%%%%%
% Fig. 7
% 
%%%%%%%%%%%%%%%%%%%%%%%%%%%%%%%%%%%%%%%%%%%%%%%%%%%%%%%%%%%%%%%%%%

\subsection{Linking GBT and VLA Studies of the IRC+10216
  Astrosphere}\protect\label{VLAassess}
As previously noted,
several of the small \HI\ clumps previously
detected with the VLA by MR07 align closely with the FUV-emitting astrosphere
of IRC+10216 (Figure~\ref{fig:vlaoverlay}). These features are now
also seen to overlap with the \HI\ shell uncovered with the
GBT (Figure~\ref{fig:rings}). However,
confirming an association between the VLA-detected ``arcs'' 
in the the northwest quadrant of Figure~\ref{fig:vlaoverlay} and the CSE and/or 
vortical tail of IRC+10216 is less straightforward.

MR07 showed that the velocity distribution of the gas comprising the VLA arcs
is consistent with a circumstellar origin. 
Matthews et al. (2011) also drew attention to what appears to be  reflective symmetry
between the \HI\ arcs and morphologically similar 
FUV features in the southern part of the vortical tail. 
The authors speculated that the \HI\ arcs might therefore
represent a portion of the tail where the \HI\ column density
is enhanced because of preferential dissociation of molecular
gas (e.g., due to an anisotropy in the UV radiation field).

As seen from a comparison of Figure~\ref{fig:vlaoverlay} with the GBT
channel maps in Figure~\ref{fig:cmaps},
the VLA arcs lie near the southeastern rim of an \HI\ cloud complex 
that extends across the northern portion of the channel maps at
velocities close to the systemic velocity of IRC+10216 (see
\S\ref{IVG}). Within this cloud complex
are  regions of enhanced column density, including a
compact ``core'' (unresolved by the GBT) lying in the vicinity of the VLA arcs. 
Based on an  \HI\ total
intensity map computed from GBT data spanning LSR velocities
from $-33$~\kms\ to $-17$~\kms\ (the same range used to compute the
VLA total intensity contours by MR07), the
peak column density  in the core is found to lie
at $\alpha_{\rm J2000}=09^{\rm h} 47^{\rm m} 20.3^{\rm s}$, $\delta_{\rm
  J2000}=13^{\circ} 32' 31.9''$. This position is near the northernmost ``lobe'' of
the VLA arc structures in Figure~\ref{fig:vlaoverlay}, 
but is $\sim$\am{4}{3} northwest of where the
peak column density in the arcs is measured. 

The core seen with the GBT is embedded in a more
extended envelope that extends southward to overlap with the position
of the VLA arcs (and with the northern edge of the \HI\
shell; see Figure~\ref{fig:rings}); however, the column density varies
smoothly across the location of the VLA
arcs themselves. Thus it is possible
that the VLA has filtered out much of the larger-scale
emission along the line-of-sight to reveal discrete emission structures
related to the IRC+10216 CSE and/or its wake that are obscured by
confusion in the GBT map. That the
curvature of the arcs follows the IRC+10216 astrosphere
rather than the edge of the cloud seen in the GBT map is also consistent with this
possibility. On the other hand, the gas velocities
associated with the arcs are predominantly blueshifted relative to the
stellar systemic velocity, in contrast to what is expected for material
in a trailing wake or a circumstellar shell, where drag from
the ISM tends to shift velocities toward zero LSR velocity. 
We conclude that an association between the VLA arcs and
the astrosphere of IRC+10216 remains plausible, although we cannot
unambiguously rule out line-of-sight confusion and are unable
to explicitly identify a counterpart to the  arcs in
the GBT data. 

\subsection{Is There \HI\ Emission Associated with the 
``Vortical Tail''?}
\begin{figure}
\includegraphics[scale=0.48]{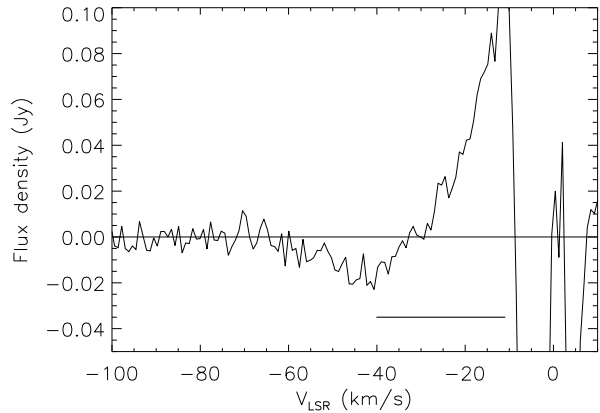}
\caption{Position-switched spectrum  obtained by differencing a
 spectrum integrated over the southern portion of the FUV-emitting vortical tail of
 IRC+10216 with a reference spectrum centered at a position $500''$ to the
 west. See \S\ref{vortical} for
 details. Across the velocity range expected for the IRC+10216 CSE
 (indicated by the horizontal bar) the line-of-sight emission
 exhibits a steep gradient with position, but there is no unambiguous evidence of
 emission associated with the IRC+10216 tail. Strong line-of-sight confusion
 at velocities $V_{\rm LSR}\gsim-18$~\kms\ precludes identification of
 a possible counterpart to the tail at higher velocities.
 }
\label{fig:tailspec}
\end{figure}

\protect\label{vortical}
We have searched our GBT data for
\HI\ emission at other locations within the FUV-defined vortical
tail of IRC+10216, outside of where \HI\ emission features were detected with
the VLA. 
Figure~\ref{fig:tailspec} shows
a position-switched spectrum, constructed by integrating
over a $310''\times450''$ box, centered at
$\alpha_{\rm J2000}=09^{\rm h} 46^{\rm m} 2.2^{\rm s}$, $\delta_{\rm
  J2000}=13^{\circ} 19' 33.0''$, and differencing with a comparable ``off''
spectrum offset $500''$ 
to the west. Gradients in the line-of-sight emission between these two
locations result in a residual slope across the velocities of interest
in the resulting difference spectrum, but we see
no compelling evidence for a distinct component of \HI\ emission associated with
the tail. 
Based on this spectrum, we place a 3$\sigma$ upper limit on the \HI\
mass in the tail over the velocity range $-40\le V_{\rm LSR}\le
-18$~\kms\ of $M_{\rm HI}<9.8\times10^{-3}~M_{\odot}$.

The lack of an \HI\ counterpart to the vortical tail 
seems somewhat surprising given that the tail lies
exterior to the \HI\ shell and therefore should be susceptible to
molecular dissociation by
the interstellar radiation field. One possibility is that gas in the
tail has been decelerated  as
a result of its interaction with the ISM (see Matthews et al. 2008,
2011) and thus would emit at
velocities shifted further toward zero LSR velocity compared with gas
in the shell. In this case, \HI\ emission redward of $V_{\rm
  LSR}\gsim-18$~\kms\ and near the location of the vortical tail
would become impossible to disentangle from the background
emission in our present data. 
Another possibility is that at least some of the
FUV emission that appears associated with the tail instead lies in the
foreground or background and is unrelated to the star.
 
Our GBT
data provide insight into another question concerning the wake of
debris trailing IRC+10216.
An inspection of the {\it GALEX} image shown in
Figure~\ref{fig:rings} reveals what appears to be a filament
connecting the vortical tail of IRC+10216 to a spatially extended
complex of FUV emission to the southwest of the star. Because this
emission lies downstream from the direction of IRC+10216's space
motion (see Table~1), this raises the intriguing possibility that
the FUV-emitting material might correspond to debris from a previous
mass-loss phase of IRC+10216. However, as we show in \S\ref{LLCC}, the
complex of FUV emission to the southwest is correlated
with enhanced \HI\ emission at a velocity of 
$V_{\rm LSR}\approx+3.5$~\kms\ that can be
attributed to the Local Leo Cold Cloud and therefore appears to be
unrelated to IRC+10216.

%%%%%%%%%%%%%%%%%%%%%%%%%%%%%%%%%%%%%%%%%%%%%%%%%%%%%%%%%%%%%%%%%%
% Fig. 8
% 
%%%%%%%%%%%%%%%%%%%%%%%%%%%%%%%%%%%%%%%%%%%%%%%%%%%%%%%%%%%%%%%%%%

%%%%%%%%%%%%%%%%%%%%%%%%%%%%%%%%%%%%%%%%%%%%%%%%%%%%%%%%%%%%%%%%%%
% Fig. 9
% 
\begin{figure}
\includegraphics[scale=0.4]{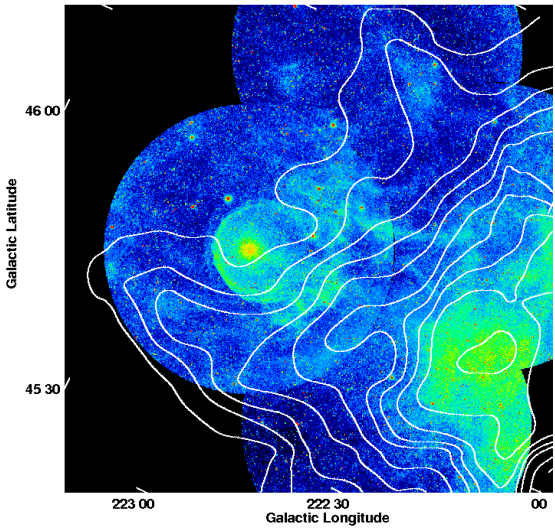}
\caption{{\it GALEX} FUV image of IRC+10216 with GBT \HI\ contours
 at $V_{\rm LSR}=3.5$~\kms\ overplotted. \HI\ emission at this
 velocity is dominated by the LLCC (see
 \S\ref{LLCC}). Contour levels are
 (1,2,3...7.5,8,8.5,9.5,10)$\times$1.2~Jy
 beam$^{-1}$.  A region of enhanced FUV surface brightness southwest of IRC+10216
 coincides with the highest \HI\ column density at this velocity,
 implying that this material is linked with the LLCC and is not 
 debris shed by IRC+10216. A Galactic coordinate system is used to
 illustrate the orientation of the LLCC contours
 relative to lines of constant Galactic latitude. }
\label{fig:LLCC}
\end{figure}
%%%%%%%%%%%%%%%%%%%%%%%%%%%%%%%%%%%%%%%%%%%%%%%%%%%%%%%%%%%%%%%%%%

%%%%%%%%%%%%%%%%%%%%%%%%%%%%%%%%%%%%%%%%%%%%%%%%%%%%%%%%%%%%%%%%%%%%%%%%%%%%%%%%%%%

\section{Discussion}\protect\label{discussion}
We have reported the discovery of a low surface brightness \HI\
shell associated with the astrosphere of IRC+10216.
Several lines of evidence support an association between the \HI\
shell and
the IRC+10216 astrosphere rather than a chance
superposition of emission along the line-of-sight: (1) the \HI\
shell matches closely in both size and position
with the FUV-emitting shell
previously detected with {\it GALEX} (see
Figure~\ref{fig:rings}); (2) the central star is offset by
$\sim110''$ relative to
 the centre of the \HI\ shell, in the direction expected for a star
 moving through the ISM (see below); (3) the narrow \HI\ linewidth of the
 shell  relative to the wind outflow speed (8.1~\kms\ compared to
 14.6~\kms) and the small offset in velocity of the peak
 shell emission toward zero LSR velocity compared with the stellar
 systemic velocity ($-$23.8~\kms\ versus $-$25.5~\kms)  are both consistent with
 \HI\ shells detected around other stars, and can be explained by the
 interaction between the shell and the surrounding ISM (G\'erard et
 al. 2011; Le~Bertre et al. 2012; Matthews et al. 2013); (4) we observe
 an 
 enhancement in the \HI\ column density along the leading edge of the
 shell, where gas pile-up is expected (Figure~4); (5) we
 detect a region of enhanced velocity dispersion coincident with the
 location of the IRC+10216 CSE (\S\ref{dispersion}).
In the sections that follow, we comment on the possible origin of the
detected \HI\ shell and place constraints on its age and on the 
fraction of the total CSE mass that is comprised of \HI\ gas.

\subsection{The Origin of the Atomic Hydrogen in the Circumstellar
  Environment of IRC+10216}
Our new observations are consistent with past indications that the bulk of
the hydrogen gas in the circumstellar environment of IRC+10216 is in
molecular form. Nonetheless, while the fraction of atomic hydrogen in
the CSE is
small, understanding the origin of this material provides important
insights into the atmospheric physics of IRC+10216, as well
as on the interaction between this mass-losing star and
its interstellar environment.

If the effective temperature of IRC+10216
is sufficiently warm ($T_{\rm eff}\sim$2200-2300~K), then model
atmospheres predict that some small
fraction of its wind ($\lsim$1\%) may  be in atomic form within a few
stellar radii owing to freeze-out---i.e., the cessation at some
critical density and pressure of the
three-body recombination needed to create H$_{2}$
(GH83).  As described in \S\ref{windHI}, our upper limit on the 
mass of \HI\ directly along the
line-of-sight to IRC+10216 is consistent with a photospheric
freeze-out abundance of \HI\ of $\lsim$0.2\%.  However, while even such
a small fraction of atomic material in the wind may produce an
observable signal when integrated over a sufficient volume, the observed
radial distribution of \HI\ column density for the case of a steady, spherically
symmetric wind scales as the inverse of the projected distance from
the star, resulting in a centrally peaked distribution (e.g., Bowers \& Knapp 1988). 
This is in contrast to the shell with a central hole that we observe
surrounding IRC+10216. This implies that some other mechanism to
produce \HI\ is
needed to fully explain the current observations. 

We note that the low observed \HI\ fraction of the IRC+10216
wind appears to be inconsistent with the stellar effective temperature of
2500-3000~K advocated by Men'shchikov et al. (2001).
Such a high value for the stellar effective temperature would imply
that hydrogen is mostly in atomic form in the star's atmosphere
(GH83). As the H/H$_{2}$ abundance ratio should be frozen out in the expanding wind,
we therefore should have easily detected the \HI\ line with an intensity $\geq$100
larger than observed. This argument is supported by the detection of
the \HI\ line in a number of stars with
effective temperatures larger than 2500~K (G\'erard \& Le~Bertre 2006,
Matthews et al. 2013). The latter detections also suggest that atomic hydrogen does not in
general recombine 
efficiently into molecular hydrogen within the outflows of these
warmer stars.

One plausible candidate for the origin of the \HI\ surrounding
IRC+10216 is
dissociation of H$_{2}$ molecules by the interstellar radiation field.
Adopting estimates for the expected strength of the
interstellar radiation field and the efficiency of the absorption of
UV photons in the Lyman and Werner bands of H$_{2}$, Morris \& Jura
(1983) provided a simple analytic formula to predict the total number of
hydrogen atoms in a circumstellar shell expected to arise from
photodissociation: $n_{\rm H}\approx 1.8\times10^{6}r^{3}/V_{\rm out}$. Taking
$r=$0.4~pc and converting the result to solar mass units, we
find $M_{\rm HI}\sim 1.9\times10^{-3}~M_{\odot}$, in agreement, to
within uncertainties, with
our measurement (\S\ref{windHI}). The radial column density profile in
this case should also peak in the outskirts of the CSE, consistent with our data.

Of course these agreements with the predictions from Morris \& Jura (1983)
must be treated cautiously since their
formula relies on a number of simplifying assumptions. These include
a particular (isotropic) strength for the Galactic radiation field and
a constant, isotropic mass-loss rate for the entire age of the
CSE. However, the interstellar 
radiation field unlikely to be uniform for a star well out of the Galactic
plane, while the assumption of constant, isotropic mass loss  
is also questionable for IRC+10216, 
given the complex shells and other structures seen in the
inner portions of the CSE at various wavelengths, suggesting that the recent 
mass-loss rate has been variable and punctuated by discrete episodes
of enhanced mass loss. Further, some of the mass
loss events may have have been non-isotropic (see references given in
\S\ref{introduction}). Nonetheless, in principle, it may be realistic
to assume that the total mass lost
during these discrete events may be small compared with the mean mass-loss rate
over the past $\sim10^{5}$~yr (e.g., Figure~8 of Vassiliadis \& Wood
1993; see also \S\ref{hydro}). 

One other key factor that needs to be considered is that the
model of Morris \& Jura (1983) assumes that the stellar wind is in
free expansion throughout the CSE and does not take into account the motion of
the star through the ISM. In general, such motion can dramatically
affect the shape, structure, density, and kinematics of the CSE, as well as 
its chemistry. For example, the motion is expected to introduce
axisymmetries in the CSE, and the star is expected to become offset from
the CSE centre (Villaver et al. 2012). Size segregation of
dust grains may also result (van Marle et al. 2011). 

IRC+10216 is known to be
moving through the ISM with a moderately high space velocity (Table~1),
and the bow shock and vortical tail structures seen in FIR and FUV
images are clear manifestations of this interaction. Interaction with
the ISM may also account for the observation that the \HI\ shell
detected with the GBT is primarily seen at LSR velocities higher 
than the systemic velocity. This type of velocity offset toward
$V_{\rm LSR}$=0~\kms\ is a well-established hallmark of ISM
interaction and has now been seen in a number of other
\HI-detected AGB stars (G\'erard et
al. 2011). Finally, the ISM interaction naturally explains the offset
(by $\sim110''$) of the
star relative to the centre of the \HI\ shell (see Figure~\ref{fig:rings}).

Another prediction of hydrodynamical models
is that the interaction with the ISM may lead to significant
quantities of interstellar hydrogen being swept into the circumstellar
environment (Villaver et al. 2002, 2012). This, for example, may
account for the arc of \HI\ emission we detect along the leading edge
of the shell, near the bow shock region, and which exhibits a larger
spread of gas velocities compared with the other portions of the shell
(see Figure~\ref{fig:cmaps} and \S\ref{windHI}). An analogous
phenomenon is seen in the case of solar-like stars moving through 
the ISM, leading to the
formation of a so-called ``hydrogen wall'', just interior to the bow shock (Wood
et al. 1996). In the case of IRC+10216, to account for the
observed mass of \HI\ within a volume of radius $r$=0.4~pc purely from
swept-up interstellar material would
require an ambient \HI\ density of $n_{\rm HI}\sim0.078$ atoms cm$^{-3}$, a
value that is quite plausible even for a star well out of the Galactic
Plane. Further, if we assume in the strong shock approximation 
that the velocity of the interstellar gas
is reduced by a factor of 4 when it crosses the bow shock, ISM
gas streaming at 42~\kms\ in the rest frame of the star would be
reduced to a velocity of $\sim$10~\kms, or 6~\kms\ when projected along the
line-of-sight. Thus the predicted LSR gas velocities in the shell might be
expected to lie roughly in the range between $-19$ to $-25$~\kms,
consistent with what we observe. 

One additional possibility for producing atomic hydrogen in the IRC+10216 CSE
may be related to the same emission mechanism responsible for the FUV
emission from the astrosphere.
The projected overlap between the \HI\ shell that we have
detected with the GBT and the FUV-emitting shell detected by
{\it GALEX} suggests that the gas giving rise to the two structures
could be co-spatial. 
Sahai \& Chronopoulos (2010) proposed that the FUV luminosity of
IRC+10216 is likely to be due to collisional excitation of H$_{2}$
molecules by hot ($\sim$30~eV) electrons, analogous to the mechanism that Martin
et al. (2007) proposed to explain the FUV-emitting wake of
Mira. In this model, the
dissociation of H$_{2}$ molecules is expected to be a byproduct of
the excitation process (Raymond et al. 1997) and should result in the
steady production of hydrogen atoms. Assuming a dissociation rate of
$\sim10^{42}$ s$^{-1}$ (a third the rate assumed for Mira;
see Martin et al.) implies that the entire observed mass of \HI\ in
the IRC+10216 could
be produced in $\sim$60,000 yr. This assumes that the formation of new
molecules would not
occur efficiently in the outer CSE because of the low gas densities,
the absence of dust grains at these radii 
(Decin et al. 2011), and the decreased shielding from the
interstellar radiation field. However, this mechanism for producing \HI\ 
also assumes that there will be a
source for the hot electrons in the astrosphere. 
The stellar space velocity assumed by Sahai \& Chronopoulos (91~\kms)
is higher than the recent determination by Menten et al. (2012), and
consequently, the temperature that we estimate for post-shock gas in the bow shock
region ($\approx$52,000~K; see \S\ref{windHI}) seems too low to
produce the $\sim$30~eV electrons required.

\subsection{Age Dating the CSE}\protect\label{hydro}
One of the hydrodynamical models presented by Villaver et al. (2012)
represents a star quite similar to IRC+10216, having
initial mass of 3.5$M_{\odot}$ and moving with a space velocity of
50~\kms\ through an interstellar medium with a
particle density of 0.1~cm$^{-3}$. Examining Figure~12 of Villaver et al. at a time
$t\sim (3.3-3.6)\times10^{5}$~yr since the start of the AGB, we note rather good
agreement  between the predicted size of the CSE and what is observed
for IRC+10216. Further, the Villaver et al. model predicts that a nearly
spherical shell of
enhanced hydrogen density should circumscribe the entire CSE, consistent
with our GBT results and with the FUV structures seen by {\it GALEX}
east of the star. 

The age of the IRC+10216 astrosphere implied by the
comparison with the
Villaver et al. model predictions is roughly an order of magnitude 
larger than the age of
the shell computed from its dynamical crossing time
($\sim$27,000~yr, assuming $r\approx$0.4~pc and $V_{\rm
  outflow}$=14.6~\kms) and roughly five times larger than 
the value of $\sim$69,000~years derived by
Sahai \& Chronopoulos (2010) based on assumptions about the
expansion timescales for the shocked and unshocked wind regions. 
This underscores the importance of taking into
account the effects of interaction with the surrounding ISM when age-dating
circumstellar ejecta (see also Matthews et al. 2008, 2011).

\subsection{Comparison of the Predicted Versus the Observed CSE Mass}
Assuming the age of the IRC+10216 CSE to be $t\sim3.5\times10^{5}$~yr
(see \S\ref{hydro}), we can readily infer that
the mass of atomic hydrogen that we observe to be associated with IRC+10216 ($M_{\rm
  HI}\sim3\times10^{-3}~M_{\odot}$)  is only a
tiny fraction of the expected total mass of the CSE.  For example,
assuming that IRC+10216 has been losing
  mass at its current high rate for $\sim$60,000~yr and at a lower
  rate prior to this (see Figure~2 of
  Villaver et al. 2012), the CSE mass should be
  $\gsim1.2~M_{\odot}$.

FIR observations also point to a significantly larger CSE mass than we have
measured from \HI. Based on 100$\mu$m observations
from {\it Herschel}, Decin et al. (2011) estimated a
mass for the IRC+10216 CSE within a radius $r<390''$ of
$\sim0.17M_{\odot}$ (scaled to our adopted distance), assuming a
gas-to-dust ratio of 250 (which is likely lower than the true value,
making the discrepancy even more severe; see,
  e.g., Groenewegen et al. 1998).  
Our observations therefore support earlier assertions, both
theoretical and observational, that most of the mass loss from
IRC+10216 has been in molecular form (e.g., Zuckerman et al. 1980;
Bowers \& Knapp 1987) and that dust and molecular
self-shielding are able to preserve a large fraction of molecular gas
even at large distances from the star $r>10^{17}$~cm (Morris \& Jura
1983; GH83). 

For completeness, we note that there is another type of effect that
in principle may lead to an
underestimate of the amount of \HI\ measured in very cool CSEs, namely 
self-absorption and/or absorption of the background emission (e.g.,
Levinson \& Brown 1980). Based on
Reich \& Reich (1986), the background temperature of the continuum
near IRC+10216 is $\sim$3.4~K, while from Figure~\ref{fig:onstar}, the
\HI\ background over the velocity range of the circumstellar shell is
$\sim$0.1~Jy (0.2~K). Thus the total background is $\le$3.6~K, and for
absorption to be occurring in the shell gas, its temperature must be
lower than this. While such low temperatures have been reported in the
circumstellar environments of some AGB and post-AGB sources 
(Sahai 1990; Sahai \& Nyman 1997), they are expected to be achieved
only in a freely expanding wind region where cooling is efficient, 
not within the zone where
circumstellar matter is slowed by the ISM (e.g., Villaver et
al. 2002), 
nor in the region inside the
bow shock, where the pile-up of interstellar matter may
occur. Absorption effects are thus likely to affect our \HI\ mass
estimate for IRC+10216 
only directly along the line-of-sight to the star (where the telescope
beam samples gas close to the star that is 
still in free expansion) and only if the atomic gas
fraction in the wind as it leaves the star is larger than predicted by
current models for  stars with $T_{\rm
  eff}\lsim$2500~K (see \S\ref{introduction}). 
However, in this case, the line profile shape
would be expected to be rectangular or parabolic,
depending on its extent relative to the telescope beam (Zuckerman
1987), in contrast to what is observed toward the stellar position
(Figure~\ref{fig:positionswitched}).

\section{Summary}
Using sensitive \HI\ mapping observations obtained with the GBT, we have
uncovered evidence for an \HI\ shell of radius $\sim$0.4~pc
and a total \HI\ mass $\sim(3.2\pm1.6)\times10^{-3}M_{\odot}$
surrounding the carbon star IRC+10216. 
The \HI\ shell  is comparable in position and scale to the star's
previously discovered FUV-emitting astrosphere (Sahai \&
Chronopoulos 2010) and encompasses some of the
\HI\ emission clumps previously seen encircling the star in \HI\
maps obtained with the VLA by MR07. 
An enhancement in \HI\ column density is seen along the
leading edge of the \HI\ shell (near the previously identified 
bow shock region), indicating a
pile-up of neutral gas at this location. We find no
evidence for atomic hydrogen associated with the wide-angle, FUV-emitting vortical
tail that trails the motion of the star through the ISM, although
line-of-sight confusion could have precluded detection if gas in the
tail has been significantly decelerated as a result of interaction with the ISM.

The quantity of atomic
hydrogen that we observe 
in the circumstellar environment of IRC+10216 is less than 1\% of 
the total predicted mass of
the CSE and is comparable to quantities of atomic hydrogen expected to
originate from
photodissociation of a predominantly molecular wind or from matter
swept from the surrounding interstellar medium. Assuming that the FUV
emission from the astrosphere arises from collisional excitation of
H$_{2}$ molecules by hot electrons,  the resulting collisional 
dissociation of molecules may also contribute to the observed \HI\
mass. Finally, a small fraction of the observed atomic hydrogen may
result from  
freeze-out in the photosphere, but the observed radial column density distribution
of \HI\ is inconsistent with this being the sole origin of the gas.  

The \HI\ shell that we detect surrounding IRC+10216 may be a precursor
of the large atomic gas shells reported previously around some planetary
nebulae (Taylor et al. 1989; Rodr\'\i guez et al. 2002) and that are
predicted by hydrodynamic models to occur during the evolution of
mass-losing AGB stars moving
through the ISM (Villaver et al. 2002, 2012).
The
detection of \HI\ associated with the CSE of IRC+10216
demonstrates the feasibility of mapping the kinematics of the
CSEs of other evolved stars with low
stellar effective temperature ($T_{\rm eff}\lsim$2500~K) 
despite the fact that their winds may be predominantly molecular.

\section*{Acknowledgments}
LDM would like to thank J. Lockman and other members of 
the NRAO Green Bank staff for
guidance in the acquisition and reduction of the GBT data presented
here and 
E. Greisen for new developments in AIPS
that aided in the analysis of these data.
These observations were taken as part of NRAO program
AGBT11B$\underline~$013. 
LDM also gratefully acknowledges financial support from grant
AST-1310930 from the
National Science Foundation.

%%%%%%%%%%%%%%%%%%%%%%%%%%%%%%%%%%%%%%%%%%%%%%%%%%%%%%%%%%%
\appendix
\section{Characterizing the ISM in the Direction of 
IRC+10216}\protect\label{ISM}
Frequency-switched \HI\ spectra toward the direction of
IRC+10216 are complex, comprising a blend of
multiple velocity components (e.g., Hartmann \& Burton 1997). 
The characterization of these components is
important for identifying and interpreting circumstellar
emission and also provides insight into the local
interstellar environment of the star. 
In this Appendix, we provide a brief description of the main
spectral components present along the line-of-sight to
IRC+10216 
based on the analysis of our new GBT data, coupled with information
from previously published studies. 

Figure~\ref{fig:onstar} (already discussed in \S\ref{results}) shows a
frequency-switched \HI\ spectrum toward the direction of IRC+10216.
Using the AIPS task {\sc XGAUS} (Greisen 2014), we have performed a
Gaussian decomposition of the line profile and find that a minimum of six emission 
components is required to adequately reproduce the spectrum. The results of the
best-fitting decomposition (as determined by  $\chi^{2}$ minimization) are
summarized in Table~A1. 

Given the large number of components in the fit, the decomposition presented in
Table~A1 may not be unique. We also ignore the possibility of 
absorption components and/or optical depth effects that might result
in non-Gaussian emission profiles
(see Verschuur \& Knapp 1971; Haud \& Kalberla 2006).
However, despite these caveats, we find these fit results are instructive
for guiding the discussion that follows. 

%%%%%TABLE A1%%%%%%%%%%%%%%%%%%%%%%%%%%%%%%%%%%%%%%%%%%%%%%%%%%%
\begin{table}
\begin{scriptsize}
\caption{Gaussian Decomposition of \HI\ Spectrum toward IRC+10216}
\begin{tabular}{llll}
\hline
Component & Amplitude (mJy) & $V_{\rm LSR}$ (\kms) & FWHM (\kms) \\
\hline

1 & 72$\pm$1 & $-29.0\pm0.2$ & 19.5$\pm$0.5\\

2 & 160$\pm$8 & $-5.82\pm0.04$ & 3.7$\pm$0.1\\

3 & 668$\pm$8 & $-4.66\pm0.05$ & 16.5$\pm$0.1\\

4 & 185$\pm$8 & $-1.0\pm$0.1 & 7.6$\pm$0.4\\

5 & 292$\pm$4 & +3.17$\pm$0.01 & 1.66$\pm$0.03\\

6 & 19$\pm$1 & +7.9$\pm$1.6 & 92.0$\pm$2.8\\

\hline
\end{tabular}

{Fit components and their uncertainties 
were determined using the AIPS task {\sc XGAUS}. 
The fitted spectrum was integrated over a
  300$''\times300''$ region centered on the position of
  IRC+10216. The RMS residual was 2.6~mJy 
over the fitted LSR velocity range from $-75$ to 87~\kms. }

\end{scriptsize}
\end{table}
%%%%%%%%%%%%%%%%%%%%%%%%%%%%%%%%%%%%%%%%%%%%%%%%%%%%%%%%%%%%%%%%%%

\subsection{ISM Local to IRC+10216?}
In terms of peak column density, 
the dominant spectral component in Figure~\ref{fig:onstar} is centered near 
$V_{\rm LSR}\approx -4.7$~\kms\ (viz. Component~3 in Table~A1). Two other weaker
components (2 and 4 in Table~A1) are also
found at similar velocities ($-5.8$~\kms\ and $-1.0$~\kms,
respectively). Toward the Galactic latitude and longitude 
of IRC+10216, all three of these components have 
velocities consistent with values expected for gas in
regular Galactic rotation (Hartmann
\& Burton 1997). This implies that some of this gas may be local to
the star or lie
within its Galactic neighborhood. However, this remains unconfirmed,
and to our knowledge,
no interstellar absorption features
have been unambiguously detected at 
comparable velocities in the spectra of background stars in this region
(Kendall et al. 2002; Mauron \& Huggins 2010).

\subsection{The Local Leo Cold Cloud (LLCC)}\protect\label{LLCC}
A second dominant component in our GBT spectrum (Component~5 in
Table~A1) is centered near
$V_{\rm LSR}\approx +3$~\kms\ and has a rather narrow linewidth 
(FWHM$\approx$1.7~\kms), indicating the presence of very cold gas. 
We attribute this component to the well-known
Local Leo Cold Cloud (LLCC). 
Portions of the LLCC were first discovered by Verschuur (1969), and
Haud (2010) later showed that the emission detected by Verschuur is
actually part of a vast ribbon of clouds of
similar linewidths and velocities, stretching from approximately 7~hr
to 11~hr in RA and $-10^{\circ}$ to
$+40^{\circ}$ in DEC. Using stellar absorption line
measurements, Peek et al. (2011b) constrained
the distance of the LLCC to lie in the range 11.3~pc to
24.3~pc---i.e., in the foreground of IRC+10216 and comfortably 
inside the Local Bubble. 
Peek et al. also demonstrated a clear correlation between
the \HI\ emission from the  LLCC and 100$\mu$m dust emission. 

Within the region we have mapped with the GBT, we find that
the highest LLCC \HI\ column densities
correlate with a region of enhanced FUV surface brightness to the
southwest of the IRC+10216 astrosphere (Figure~\ref{fig:LLCC}). 
This correlation implies that this FUV-emitting
material is not debris shed by IRC+10216, despite its location
downstream of IRC+10216's space trajectory and in spite of what 
appear to be FUV-emitting
filaments connecting this cloud to the astrosphere. 

\subsection{High-Velocity Halo Gas?}
Broad, high-velocity wings are another feature of the spectrum shown in
Figure~\ref{fig:onstar}. These wings 
are also seen in the GALFA spectrum presented by 
Menten et al. (2012), confirming they are not artefacts caused by poor
baseline fits or erroneous stray radiation corrections. 
At positive LSR velocities,
low-level emission can be seen stretching to velocities as high as
$\sim$100~\kms, and we were able to parametrize this with a broad Gaussian of
FWHM$\sim$91.5~\kms, centered at $V_{\rm LSR}=7.9$~\kms\ (Component~6
in Table~A1).
While it is reasonable to question whether the large gas dispersion
implied by such a broad Gaussian is physically meaningful, we note that similarly
broad Gaussian components have been reported by other workers at high
Galactic latitude and have been suggested to 
arise from halo gas with a high velocity dispersion (Kalberla et
al. 1998; Haud \& Kalberla 2006).  

Regardless of whether the extended red line wing in the spectrum 
represents a single velocity
component or a blend of several along the line-of-sight, 
the presence of gas extending to such high velocities cannot be accounted for by 
Galactic disc rotation, particularly given the relatively high Galactic
latitude of IRC+10216 (Stark et al. 1994; Hartmann \& Burton 1997).  
Indeed, based on maps presented by Wakker \& van Woerden (2013), the
IRC+10216 field lies near the outskirts of the ``WA'' high-velocity
cloud complex, which is characterized by LSR 
velocities in the range 80-195~\kms\ (Wakker \& van Woerden 1991). An
association between the high positive velocity gas we observe and the WA
complex is therefore possible. Distance estimates to the WA 
complex place this
material at 15-20~kpc (Wakker \& van Woerden 2013), or well out in the
Galactic halo.  

\subsection{Blueshifted Intermediate Velocity Gas}\protect\label{IVG}
At negative LSR velocities, a ``shoulder'' of
emission is seen blueward of $V_{\rm LSR}\lsim-20$~\kms\ in the
spectrum in Figure~\ref{fig:onstar}. 
In the fit presented in Table~A1, we have characterized this
emission with a Gaussian centered at $V_{\rm
  LSR}\approx -29$~\kms\ and a FWHM linewidth of 19.5~\kms\
(Component~1 in Table~A1). The peak column density of
this component therefore occurs at a velocity close to 
the stellar systemic velocity of
IRC+10216, and indeed, as described in \S\ref{windHI}, there is 
evidence that a small portion of this
emission is in fact related to the IRC+10216 CSE. However, as seen from our own 
channel maps of the region (Figure~\ref{fig:cmaps}), 
as well from  larger-scale maps of Galactic \HI\ (e.g., 
Hartmann \& Burton 1997), 
gas spanning these
velocities is widespread throughout the region, particularly to the
north of IRC+10216, and is far too pervasive to
be accounted for entirely from mass loss from IRC+10216. 

In optical
absorption line spectra of three stars within \ad{2}{5}
of IRC+10216, Kendall et
al. (2002) detected an absorption feature at $V_{\rm
  LSR}\approx-10$~\kms\ that they attributed to a diffuse interstellar
band. One of the targets stars lies
in the foreground of IRC+10216 (at a distance of
$\sim$55~pc), and Menten et
al. (2012) argued that gas at even higher negative velocities 
is also likely to lie in the foreground of
IRC+10216. While this cannot be excluded, Menten et al.'s arguments
are based on the assumption of monotonic Galactic rotation. However,
we note that toward the direction of IRC+10216, 
gas velocities of $|V_{\rm LSR}|\gsim$20~\kms\ are ``forbidden''
and lie outside the range nominally expected for disc
rotation (e.g., Hartmann \& Burton
1997). Thus according to the definition of Wakker \& van Woerden (2013),
the material comprising Component~1 in our spectrum
qualifies as intermediate velocity gas. 

Based on maps of the distributions of known intermediate velocity clouds
(Wakker \& van Woerden 2013), we see that IRC+10216 sightline lies in
proximity to 
the IV~Arch intermediate velocity cloud 
complex. This complex includes gas with velocities in the
range $-60 \lsim V_{\rm LSR} \lsim -27.5~\kms$ (with 
the lower negative velocity cutoff being somewhat ambiguous owing
to confusion from Galactic emission; Kuntz \& Danly 1996). These
velocities are
comparable to those represented by Component~1 of our GBT
spectrum, although based on the results of Kuntz \& Danly, 
the peak column density gas
in the IV~Arch complex more typically arises $\sim$10-15~\kms\ blueward of
the peak column density of Component~1. We conclude that an
association between Component~1 and the IV~Arch
complex is plausible but cannot be made conclusively. Based on the results
of Wakker (2001), the distance
of the IV~Arch complex lies in the range 0.8-1.8~kpc, or at several
times the distance of IRC+10216. 

%%%%%%%%%%%%%%%%%%%%%%%%%%%%%%%%%%%%%%%%%%%%%%%%%%%%%%%%%%%%%%%%%%%%%%%%

\end{document}